# Dispersive hybrid states and bandgap in zigzag Graphene/BN heterostructures

**Van-Truong Tran, Jérôme Saint-Martin, Philippe Dollfus**

IEF, Université Paris-sud, CNRS, UMR 8622, Bât 220, 91405 Orsay, France.

Emails: van-truong.tran@u-psud.fr and philippe.dollfus@u-psud.fr

**Abstract**

We study the properties of edge states in in-plane heterostructures made of adjacent zigzag graphene and BN ribbons. While in pure zigzag graphene nanoribbons, gapless edge states are nearly flat and cannot contribute significantly to the conduction, at BN/Graphene interfaces the properties of these states are significantly modified. They are still strongly localized at the zigzag edges of graphene but they exhibit a high group velocity up to $4.3\times10^5$ m/s at the B/C interface and even $7.4\times10^5$ m/s at the N-C interface. For a given wave vector the velocities of N/C and B/C hybrid interface states have opposite signs. Additionally, in the case of asymmetric structure BN/Graphene/BN, a bandgap of about 207 meV is open for sub-ribbon widths of 5 nm. These specific properties suggest new ways to engineer and control the transport properties of graphene nanostructures.

## 1. Introduction

Cutting or patterning graphene sheets into nanoribbons (GNRs) is the most common way to open a bandgap in graphene [1,2] and thus to overcome the main intrinsic limitation of this material towards many types of high performance device able to take advantage of its exceptional properties.[3–5] In the ideal case of perfect ribbons, GNRs with edges along the armchair direction (AGNRs) seem to be the most useful since their bandgap can be tuned on a wide range according to their width, which has been predicted theoretically in the early stage of graphene research.[6–8] It has been confirmed experimentally from conduction measurements [2] though the edge disorder significantly affects the band structure.[9] Actually, in the presence of edge disorder, localized states are generated and the quasi-mobility edge, or conduction gap, becomes independent on the GNR orientation, tending to follow a quasi-universal function of width.[2,9] The transport properties are of course strongly affected.[10]

For perfect GNRs oriented along the zigzag direction (ZGNRs), the actual bandgap is zero regardless of the GNR width.[6,11,12] However, a remarkable feature arises in the band structure. The top of the valence band and the bottom of the conduction band are always degenerate at $k = \pm\pi/a$. Actually these degenerate states correspond to states localized in the vicinity of the zigzag edges and their probability density decays exponentially into the center of the ribbon, which has been confirmed experimentally.[13] Additionally, it has been suggested that an opposite spin orientation between ferromagnetically ordered edge states on each edge is the ground-state spin configuration in ZGNRs, leading to possible applications for spintronics. [11,14,15] From simple nearest-neighbor tight-binding calculation, these edge states appear in the band structure as flat bands, i.e. with zero group velocity.[6,11,12,15] Calculations from more sophisticated tight-binding models and ab initio methods have shown that the bands of these states are not perfectly flat but exhibits a very small group velocity compared to the high velocity of bulk states.[10,16] Hence, they cannot reasonably be used for transport applications, though ZGNR-based devices with high negative differential conductance have been proposed [17,18] taking advantage of peculiar symmetry properties as the parity selective rule.[16]

Recently, studies of topological insulators (TI) have revealed the existence of very high group velocity in states localized at the surface or edge of samples, while bulk states have the usual properties of states in conventional insulators. This behavior has been first predicted by Kane and Mele in 2005 [19] and then demonstrated in two-dimensional (2D) HgTe/CdTe quantum wells, both theoretically [20] and experimentally.[21] TI states and topological phase transitions have been also evidenced in three-dimensional (3D) materials as $Bi_{1-x}Sb_x$ alloys, $Bi_2Se_3$ and $Bi_2Te_3$ .[22–27] Very recently, a new strategy has been proposed to generate a 2D TI state that exhibits a large bulk gap by constructing quantum well structures where a graphene layer is sandwiched between thin slabs of $Bi_2Se_3$.[28] The authors have shown that the strong hybridization between graphene and $Bi_2Se_3$ across the interfaces opens a significant energy gap (30−50 meV) and produces a 2D TI state that is driven by a new mechanism involving an inversion of the graphene Dirac bands and the conduction bands of $Bi_2Se_3$.

More recently, the strong similarity of crystallographic structure between graphene and hexagonal borin-nitride (h-BN) has been exploited new electronic properties in layered graphene-like materials, either in the form of vertical Van der Waals structures or as in-plane monolayer heterostructures. The electronic properties of the latter type of graphene/BN heterostructures have been explored theoretically [29–37] and their synthesis has been successfully demonstrated [38–41], providing additional flexibility to design new or improved devices.[42–44] Most of works were focused on the possibility of tuning the bangap, while just a few of them have paid attention to the specific electronic states at Graphene/BN interfaces.[30,36,45] In Ref. [30] Fan et al. studied the phase transition from semiconducting to half-metal and ferromagnetic metal in Graphene/BN ribbons with zigzag interface. They found out that the transition takes place when increasing the GNR width and depends strongly on whether the inferace type is B-C or N-C. In Ref. [36] Jung et al. have focused on the confinement effect induced when inserting an extremely narrowed graphene ribbon (nanoroad) in a BN sheet. They demonstrated the emergence of valley states in the midgap of BN band structure. However, this effect is strongly dependent on the width of the embedded graphene ribbon, and the physics of structures made of large graphene areas connected to BN regions was still pending. Very recently, Drost et al, [45] demonstrated experimentally and theoretically the existence of interface states in zigzag Graphene/BN heterostructures. However, this work did not investigate the details of the physical properties of these sates. In this article, we investigate the band structure of a series of zigzag graphene/BN (G/BN) heterostructures with each sub-ribbon width larger than 5nm. On the basis of second nearest-neighbor tight-binding calculations we show that a bandgap of 207 meV can be opened in a zigzag BN/G/BN structure with B-C--C-N type of bonding, even with sub-ribbons as large as 5 nm. More interestingly, we evidence the emergence of interface states that look like edge states in ZGNRs but with high group velocity, as can be observed in TIs. These interface states,are localized in the graphene side of G/BN structures. Their group velocity reaches $4.3 \times 10^5$ m/s at B-C interfaces and even $7.4 \times 10^5$ m/s at N-C interfaces. By using an effective model of single graphene ribbon, we also give a simple interpretation of the role of B(N) atoms on the formation of these interface hybrid states.

## 2. Methodology

To investigate the band structure properties of hybrid G/BN ribbons, second nearest-neighbor (2NN) tight-binding (TB) Hamiltonians have been considered, i.e. in the form [8,35]

$$H = \sum_i \varepsilon_i |i\rangle\langle i| - \sum_{\langle i,j \rangle} t_{ij} |i\rangle\langle j| - \sum_{\langle\langle i,j \rangle\rangle} t'_{ij} |i\rangle\langle j| + \sum_i V_i^{dec} |i\rangle\langle i| \qquad (1)$$

where $\varepsilon_i$ is the on-site energy at site $i$, $t_{ij}$ and $t'_{ij}$ are the hoping-parameters for nearest neighbors and next nearest neighbor interactions between sites $i$ and $j$, respectively. The potential $V_i^{dec} = P_{B(N)} \varepsilon_{B(N)} e^{-d_i/\lambda}$ is the decaying potential from B(N) atoms at site $i$ of graphene, where $d_i$ is the distance between B(N) interface atoms and the $i$-th carbon atom, $P_B$ and $P_N$ are the strength of the potential and $\lambda$ is the characteristic decay length.

To accurately describe the band structure of BN ribbons within a TB approach, it was proven that the use of a decaying potential is mandatory, in particular to remove the degeneracy at the X point.[46] However, in a pure graphene ribbon, this term is negligible because it is proportional to the onsite-energy of edge atoms, that is zero or very small for carbon atoms of both sublattices .[35,46] All TB parameters in (1) used in this paper were taken from ref [35] where they have been adjusted to provide results fitting very well with ab initio calculations in G/BN ribbons.

For infinite ribbons the Schrödinger equation can be computed by considering Block wave functions constructed as [47,48]

$$|\psi_{\vec{k}}\rangle = \sum_{\alpha=1}^{\infty} \sum_{j=1}^{N} c_j \, e^{i\vec{k}.\vec{R_\alpha}} |j_\alpha\rangle \qquad (2)$$

where $|j_\alpha\rangle$ is the $p_z$ orbital of the $j$-th atom in the unit cell $\alpha$ and $N$ is the number of atoms in a unit cell. The coefficients $c_j$ are chosen such that the wave function is an eigenstate of the tight-binding Hamiltonian. From one unit cell to another the wave function differs only by a plane wave function. Then, the Schrödinger equation for the whole periodic system can be reduced to the equation of a single unit cell, i.e. [48,49]

$$h\phi = E\phi \qquad (3)$$

$$\text{with} \quad \begin{cases} \phi = \begin{pmatrix} c_1 & c_2 & \cdots & c_N \end{pmatrix}^T \\ h = H_{n,n-1}\, e^{-ik_x a_x} + H_{n,n} + H_{n,n+1}\, e^{ik_x a_x} + V_{n,n} \end{cases}$$

where $H_{n,n}$ is the Hamiltonian of the *n-th* unit cell, $H_{n,n\pm 1}$ is the coupling term between the *n-th* and *n±1-th* unit cells, $V_{n,n}$ is the matrix of the decaying potential in the *n*–th unit cell and $a_x = \sqrt{3}\, a_0$, $a_0$ being the distance between two nearest neighbors.

Once the eigen energies $E$ and the coefficients of the eigen function $\phi$ are solved for each wave vector $k_x$, the group velocities can be numerically computed as $v_x = \dfrac{1}{\hbar} \cdot \dfrac{\partial E}{\partial k_x}$.

## 3. Results and discussion

As reported in many previous works on zigzag graphene nanoribbons [6,11,12], nearly flat bands appear in the dispersion relation near the neutral point. They correspond to edge stages, i.e. states associated to electron densities strongly localized at the edges of the graphene ribbons. In contrast, in perfect armchair GNRs all states are delocalized. Although these localized states in ZGNRs are closed to the neutral point and are thus easily accessible, they do not contribute significantly to charge transport [16] because their group velocity is weak. Hence, they are useless for standard electronic applications.

In this section, we investigate the particular states localized at the interface between graphene and BN zigzag nanoribbons within different kinds of in-plane G/BN heterostructures. Throughout this article, the notation "B-C" ("N-C") refers to interfaces where only Boron (Nitrogen) atoms are bonded with Carbon atoms. The width of each graphene and BN sub-ribbon is defined by the number of zigzag lines $M_{CC}$ and $M_{BN}$, respectively. Unless otherwise stated, each of these numbers is 50 atomic lines, corresponding to a width of about 10 nm. Such a ribbon width is already achievable with current technology [50,51].

### 3.1. Hybrid states in a "one interface" structure

We first investigate a zigzag hybrid ribbon with just a single Graphene/BN interface, as schematized in Fig. 1(a). For convenience in calculation and analysis of results, each unit cell is divided into two non-equivalent sub-layers L1 and L2, as shown in Fig. 1(a). In Fig. 1(b), we plot the profile of the decaying potential along the width of each of these sub-layers. It is assumed that the decaying potential associated with B or N atoms only affects the sites of carbon atoms within a single slice along the *y* axis and is equal to zero in BN regions.[35,46] The band structure computed for $M_{BN} = 50$ and $M_{CC} = 50$ is plotted in Fig. 1(c). The band structure is particularly interesting near the neutrality point, between the lowest conduction band and the highest valence band. In the wave vector ranges $-\pi/a_x \leq k_x < -2\pi/3a_x$ and $2\pi/3a_x < k_x \leq \pi/a_x$, we observe two relatively flat bands, while two other bands with finite group velocity extend in the negative energy region (red solid lines).

The localization of the wave functions associated with three particular states marked (1), (2) and (3) in Fig. 1(c) is illustrated in Fig. 1(d) where we plot the corresponding profiles of probability density $|\phi|^2$. On the purple curve, the probability density corresponding to the propagative state (3) of high group velocity, marked with a purple filled circle in Fig. 1(c), is shown to be distributed along the full width of the graphene ribbon. This state is thus a classical "bulk" state. In contrast, the states (2) and (1) marked in Fig. 1(c) with a cyan filled circle and a black filled circle, respectively, correspond to wave functions strongly localized at the edges of the graphene ribbon along the graphene/BN and graphene/vacuum interfaces, respectively. The edge state of type (1) has a nearly zero group velocity and corresponds to the edge state of an isolated zigzag graphene ribbon.[6,11,12] However, the edge state of type (2), though localized at the B-C interface, is strongly dispersive and exhibits a high group velocity. It should be noted that in this case the state (2') belonging to the other dispersive band at opposite wave vector $k_{2'} = -k_2$ is localized at the same interface as the state (2) but with opposite group velocity. It will be further discussed later. The inset of Fig. 1(c) schematizes the location and the propagation properties of both the zero velocity (top black line) and high velocity (solid arrow for the left band and dashed arrow for the right band) edge states. This peculiar edge state at G/BN interface can significantly contribute to charge transport. It results from the hybridization of carbon and BN states. For the sake of simplicity we call this kind of states "hybrid state". These localized interface states extend from the edges of the Brillouin zone to the points $k_x = \pm 2\pi/3a_x$. This primary conclusion on hybrid states contradicts the discussion in ref [45] wherein the whole highest valence band (or the whole lowest conduction band) was supposed to be formed by localized interface states.

## 3.2. Hybrid states in a "two interface" structure

To further analyze the G/BN hybrid states, we now investigate structures with two G/BN interfaces. At the first interface the carbon atoms are bonded with Boron atoms. Two cases are considered for the second interface. In the first case the second interface is similar to the first one, which leads to the symmetric structure referred to as B-C--C-B. In the second case, the second interface is made of C-N bonds. The mirror symmetry is broken and the structure is referred to as B-C--C-N.

### 3.2.1. B-C--C-B structure

The structure B-C--C-B is schematized in Fig. 2(a). Here the GNR is connected with Boron atoms at both interfaces. Fig. 2(b) shows the energy dispersion of this structure. The flat bands previously evidenced in B-C structure (see Fig. 1(c)) are no longer present in this B-C--C-B system. It looks as if the flat band inherent in the C-vacuum interface has been curved downward by B-C bonds, leading to two doubly degenerate dispersive bands with high velocity. As previously observed in the simple B-C system (sub-section 1), these degenerate hybrid states (1) and (2) are localized at the graphene/BN interfaces, as shown in Fig. 2(c). They can contribute symmetrically, i.e. at both interfaces, to charge transport.

### 3.2.2. B-C--C-N structure

In the B-C--C-N structure, schematized in Fig. 3(a), carbon atoms in graphene are now connected to N atoms at the second interface. This change induces new specific features in the band structure plotted in Fig. 3(b). The original flat band is curved upward, and belongs unambiguously to the conduction band. More interestingly, an energy gap is open and separates the first conduction band from the first valence band.

The analysis of the localization of states through the probability densities plotted in Fig. 3(c) clearly demonstrates that the state (1) of the conduction band, i.e. with a dispersive band in the positive energy region, is associated with the C-N interface and thus is a hybrid state resulting from C-N bonds. Actually, the upward or downward bending of the hybrid bands depends on the on-site energy of the atoms to which edge carbon atoms are attached. When they are attached to Boron atoms of positive on-site energy ($\varepsilon_B = 1.95\,\text{eV}$) the band is bent downward, towards negative energies. In contrast, it is bent upward when Carbon atoms are connected to Nitrogen atoms with negative on-site energy ($\varepsilon_N = -\varepsilon_B = -1.95\,\text{eV}$). These features can be observed separately in B-C, N-C (not shown), B-C--C-B and N-C--C-N (not shown) structures or together in the case of the mixed structure B-C--C-N. The asymmetry of the B-C and C-N hybrid states is essentially induced by the difference of the bonding energies between the two types of bond. Since the hopping energy between B and C atoms ($t_{BC}$ = 2.0 eV) is smaller than between N and C atoms ($t_{NC}$ = 2.5 eV) the curvature is smaller in the former case

In the case of symmetric systems as B-C--C-B (see Fig. 2(b)) and N-C--C-N (not shown) the gap is equal to zero, while it is very small in pure ZGNRs (not shown). Our investigations of different structures including B-C, N-C (not shown) or B-C--C-N structures indicate that the latter gives the largest energy bandgap for a given graphene ribbon width $M_{CC}$. For example, for $M_{CC}$ = 25 (i.e. for a 5 nm width), the structures B-C, N-C and B-C--C-N exhibit a bandgap of 47 meV, 167 meV and 207 meV, respectively. In Fig. 3(d), the energy gap in structure B-C--C-N is plotted as a function of the sub-ribbon width $M_{CC}$ for different BN widths $M_{BN}$. Beyond $M_{BN}$ = 2, the gap is very weakly sensitive to the BN width, while bandgaps higher than 0.4 eV are achievable for $M_{CC}$ < 10 thanks to confinement effects.

Thus, the use of zigzag graphene/BN interface not only reveals a new kind of propagative edge states but also opens an energy bandgap. This gives new opportunities for bandgap engineering of graphene and for manipulating the transport properties of graphene nanostructures and devices.

### 3.3. Dispersion of hybrid states: specific properties

As mentioned above, the energy gap depends strongly on the graphene ribbon width $M_{CC}$. It is thus relevant to analyze in more details the $M_{CC}$-dependence of the energy dispersion of hybrid states. In Fig. 4(a), the energy dispersions of the lowest conduction band and highest valence band in the B-C-C-N structure is plotted for $M_{CC}$ = 10 and $M_{CC}$ = 50. In each of these bands a specific value of wave vector $k_t$ separates the hybrid states (symbols) from the bulk states (dashed line). In each case the value of $k_t$ has been defined from the plot of the ratio of the probability density at the center of the graphene ribbon to the probability density at the ribbon edge, divided by the width $M_{CC}$ (Fig. 4(b)). This parameter is called here "bulk/edge ratio". As shown in Fig. 4(b), the separation between hybrid states (left side) and bulk states (right side) is characterized by a transition region with a quasi-linear change of the bulk/edge ratio as a function of $k$. Consistently, the value of $k_t$ has been defined in each band as the intersection of the slope of the bulk/edge ratio in the transition region with the wave vector axis. It appears that $k_t$ is dependent on the width $M_{CC}$ and tends to the value $\pm 2\pi/3a_x$ at large width, i.e. to the position of the center of Dirac cones in graphene. It is consistent with the width-dependence of the position $k_c$ of the extrema of conduction and valence bands due to the confinement effects in GNRs,[12] together with the dependence of bandgap. Beyond this behavior of the extrema of the bands, it is remarkable that the main part of the energy dispersion of edge states, i.e. towards the Brillouin zone edge $k = \pm \pi/a_x$, is almost independent on $M_{CC}$. Thus, the dispersion and the group velocity of hybrid states are stable and insensitive to the graphene ribbon width, except near the neutrality point.

In all these structures, as in topological insulators, to each edge state of positive wave vector $+k_x$ corresponds a symmetric edge state at $-k_x$ of opposite velocity. In TIs these two states are distributed on two different sides of the sample [52] while at B-C or N-C interfaces, they are both localized at the same interface. In symmetric B-C--C-B and N-C--C-N structures, the edge states are doubly degenerate and distributed on both interfaces of the graphene sub-ribbon. In the mixed asymmetric structure B-C--C-N, two different dispersive hybrid edge bands are present, one with states localized at the B-C interface, and the other with states localized at the N-C interface, which is a specific property of this system.

To quantify the potential of the hybrid states in terms of transport properties, the group velocities of the upward and downward bands of edge states in structures B-C--C-B and B-C--C-N are plotted in Fig. 5. The edge states in pure zigzag graphene ribbon (circles) have a zero group velocity while the velocity of hybrid states reaches high values at high wave vector. We can observe that the green curve (crosses) related to the lowest band in B-C--C-N is identical to the curve with black squares related to B-C--C-B, because they are associated with the same type of B-C interface. These B-C hybrid states have a maximum group velocity as high as $4.3 \times 10^5$ m/s. In the case of N-C hybrid states (red triangles), the maximum velocity even reaches $7.4 \times 10^5$ m/s.

### 3.4. Hybrid states: the role of B(N) atoms

We already mentioned that the behavior of hybrid states is mainly governed by the values of on-site energies for B and N atoms and the hopping energies of B-C and N-C bonds. This can be easily checked by using a simple 1NN TB model without any decaying potential, i.e. considering only interactions with the first nearest neighbors, as widely done in previous works on zigzag graphene ribbons.[6,11] Actually, the results obtained (not shown) are qualitatively very similar to that obtained with the 2NN model including decaying potential, from both points of view of the energy dispersion and of the localization of states at the interfaces. The resulting effect of these TB parameters applied to the edges of the graphene ribbon is to change the energies of edge states in graphene, leading to the so-called hybrid states. This effect may thus be seen as qualitatively equivalent to the effect of changing the potential on the edge atoms of graphene.

On the basis of this analysis, we propose a simple model that can generate artificially edge states with dispersive energy bands in a ZGNR, and able to mimic the emergence of hybrid states induced by B and N atoms. It consists in applying an effective potential $V_{eff}$ as new on-site energy to the edge atoms of the ribbon, as illustrated in Fig. 6(a). The 1NN approach is used for this simplified approach.

In Figs. 6(b-e), the energy band of a ZGNR is plotted for different values of the effective potential $V_{eff}$ varying from -0.1 to 0.5 eV, applied on a single side of the ribbon. While applying a negative potential leads to the appearance of a dispersive band toward negative energies (see Fig. 6(b)), a positive potential leads to a positive energy band (see Fig. 6(c)). Consistently, the slope of the curve is all the higher if the effective potential is stronger (see Figs. 6(c) to 6(e)). The probability density associated with one typical state in the dispersive band of Fig. 6(e) is plotted in Fig. 6(f). This state, as all states in the band, is actually localized at the interface. Thus, the first layer of N (with negative onsite energy) atoms at B-C interface or the first layer of B atoms (with positive onsite energy) at N-C interface of G/BN systems plays a role similar to that of edge C atoms with modified on-site energy in a GNR.

It should be noticed that within this simple 1NN model, the highest (or lowest) energy of a dispersive hybrid band (at Brillouin zone edge $k_x = \pm\pi/a_x$) is directly equal to the applied potential $V_{eff}$. Hence, by applying on each side two different effective potentials equal to 1.394 eV and -0.75 eV, respectively, i.e. the actual energy values of the hybrid states at the Brillouin zone edge in Fig. 3(b), the energy dispersions plotted in Fig. 7 are obtained (symbols solid lines). They are very close to that computed via the complete 2NN approach (including second neighbors and the decaying potential) in the B-C--C-N system. It is thus possible to mimic correctly the hybrid states in zigzag G/BN in-plane heterostructures by using a simple model of ZGNRs with appropriate on-site energies on the edge atoms. However, hybrid states cannot be found in the case of interface with a too high (absolute) onsite energy leading to negligible hybridization, as for instance with Hydrogen atoms.[8,46]

## 4. Conclusions

In this work we have investigated a new way to engineer energy bands in graphene, by taking advantage of the possibility to grow in-plane graphene/BN heterostructures. We demonstrate the presence of a new kind of hybrid states at the interface between zigzag graphene and BN ribbons. This states are strongly localized close to the interfaces as in pure ZGNRs, but they are able to propagate with a group velocity as high as $7.4 \times 10^5$ m/s. Besides, the properties of the hybrid states differ according to the type of bonding atom at the interface. For instance, the group velocity of the hybrid state is higher at the N-C interface than at the B-C one. Moreover, in the case of asymmetric B-C--C-N structure, a relatively wide bandgap of 207 meV can be open for a graphene width of 5 nm, with strong dependence on the graphene width. Finally, we have generalized our study to other systems and discussed the conditions for the appearance of hybrid state with interesting properties. Beyond the experimental evidence of these states by Drost et al,[45] using scanning tunneling microscopy (STM), their dispersive character and the resulting bandgap could be accurately observed by using angle-resolved photoemission spectroscopy (ARPES).[23,53] The synthesis of this type of graphene/BN interface may thus open a new way of modulating and controlling the electronic properties of graphene nanostructures and devices.

## Acknowledgments

This work was partially supported by the French ANR through project NOE (12JS03-006-01).

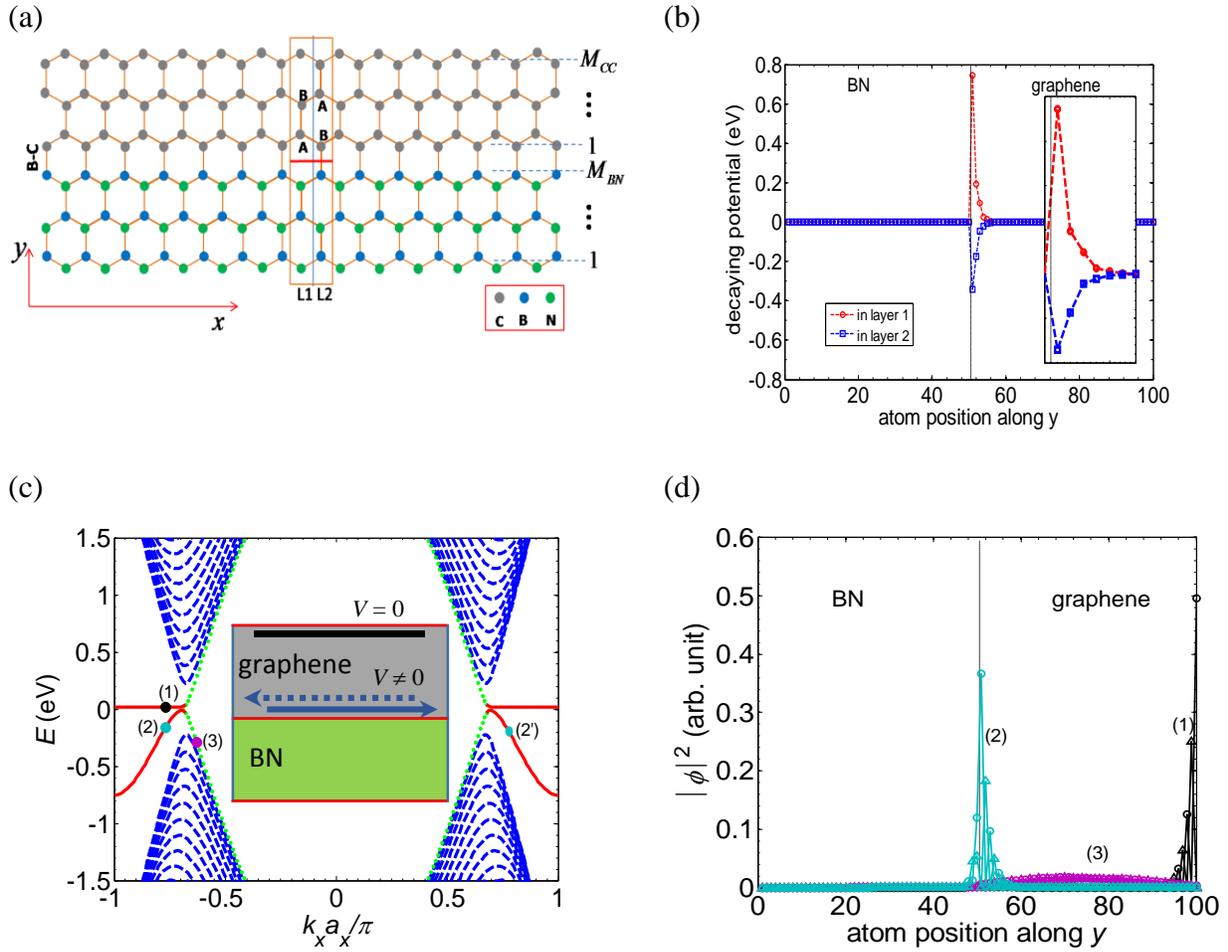

Fig 1. (a) Schematic view of zigzag Graphene/BN in-plane heterostructure with B-C bonds. The sub-ribbon widths are defined by the numbers of zigzag lines $M_{CC} = M_{BN} = 50$. (b) Decaying potential $V^{dec}$ related to B and N atoms in the layers L1 and L2 defined in (a), with BN/G boundary indicated by a vertical dashed line. Inset: magnification on the significant range of variation of $V^{dec}$ (c) Band structure, with hybrid/edge states in solid red lines. Inset: schematic representation of the velocity direction of hybrid/edge states. (d) Profile of the probability density for the three states (1), (2) and (3) defined in (c).

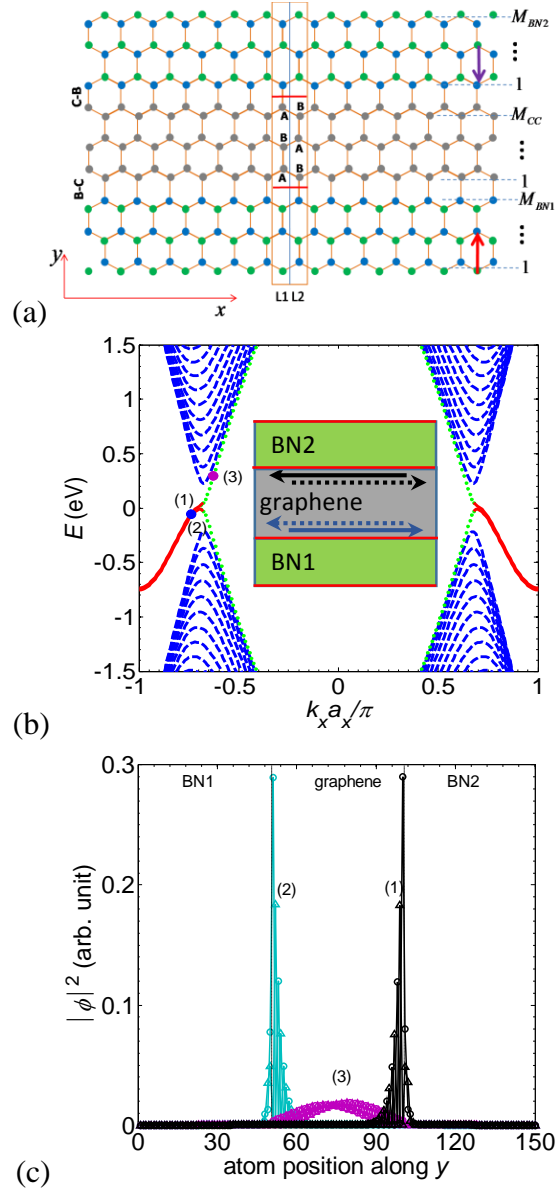

Figure 2. (a) Schematic view of a zigzag BN/G/BN structure with two C-B bond interfaces (B-C--C-B). (b) Energy band structure for $M_{BN1} = M_{CC} = M_{BN2} = 50$. Inset: schematic representation of the velocity direction of hybrid states at both interfaces (c) Profile of the probability density for the three states (1), (2) and (3) indicated in (b).

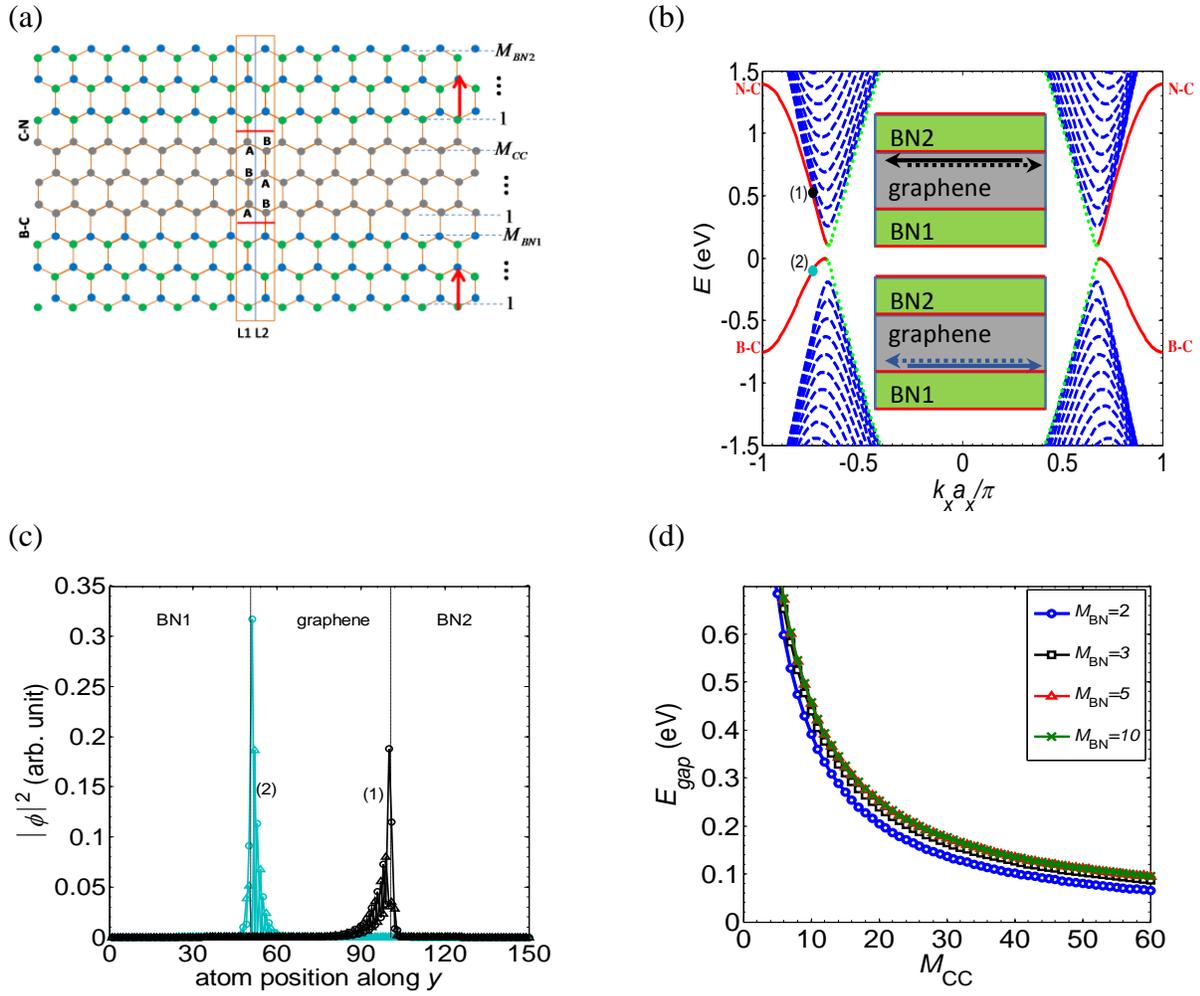

Fig 3. (a) Schematic view of a zigzag BN/G/BN with B-C--C-N bond interfaces. (b) Band structure for $M_{BN1} = M_{CC} = M_{BN2} = 50$. (c) Profile of the probability density for the two states (1) and (2) indicated in (b). (d) Bandgap as a function of sub-ribbon width $M_{CC}$ for different values of $M_{BN}$.

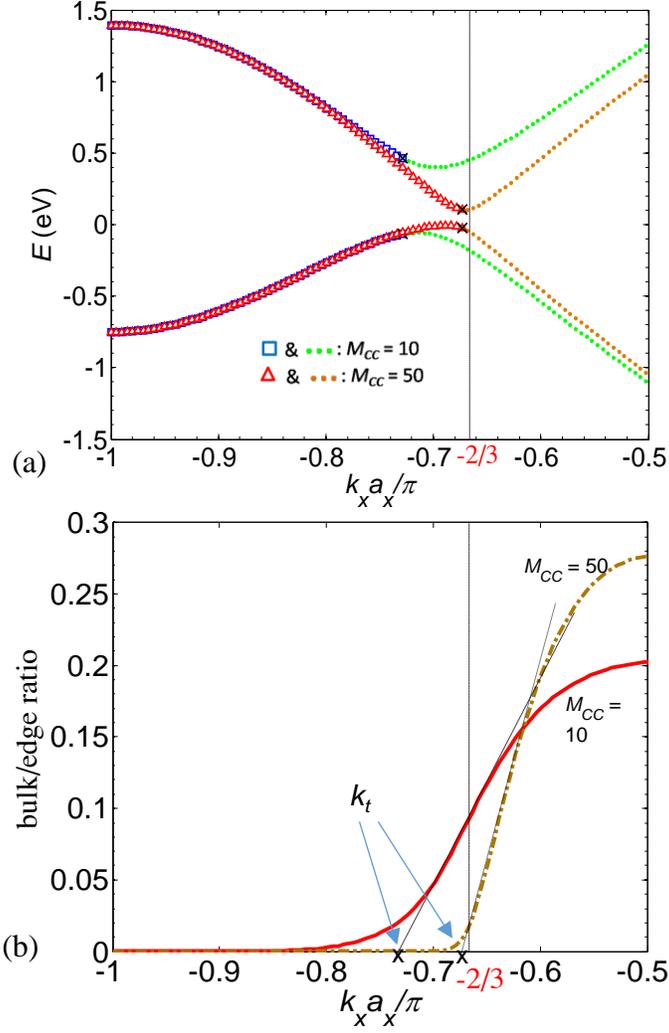

Fig 4. (a) Comparison of lowest conduction and highest valence bands in the B-C--C-N structure for different graphene ribbon width $M_{CC}$ = 10 and 50. On each band, hybrid states are marked with symbols and bulk states are in dashed line. (b) bulk/edge ratio (see text) in the highest valence band. The intersection of the slope in dashed line with the abscissa gives the value of $k_t$ separating edge states and bulk states.

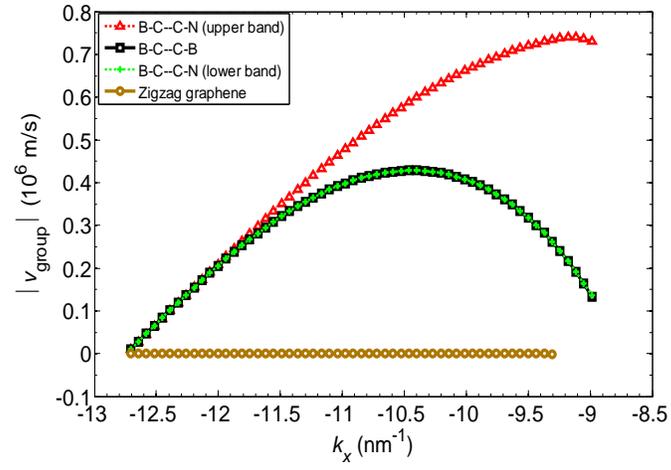

Fig 5. Group velocity of hybrid states as a function of the wave vector $k_x$ in different structures.

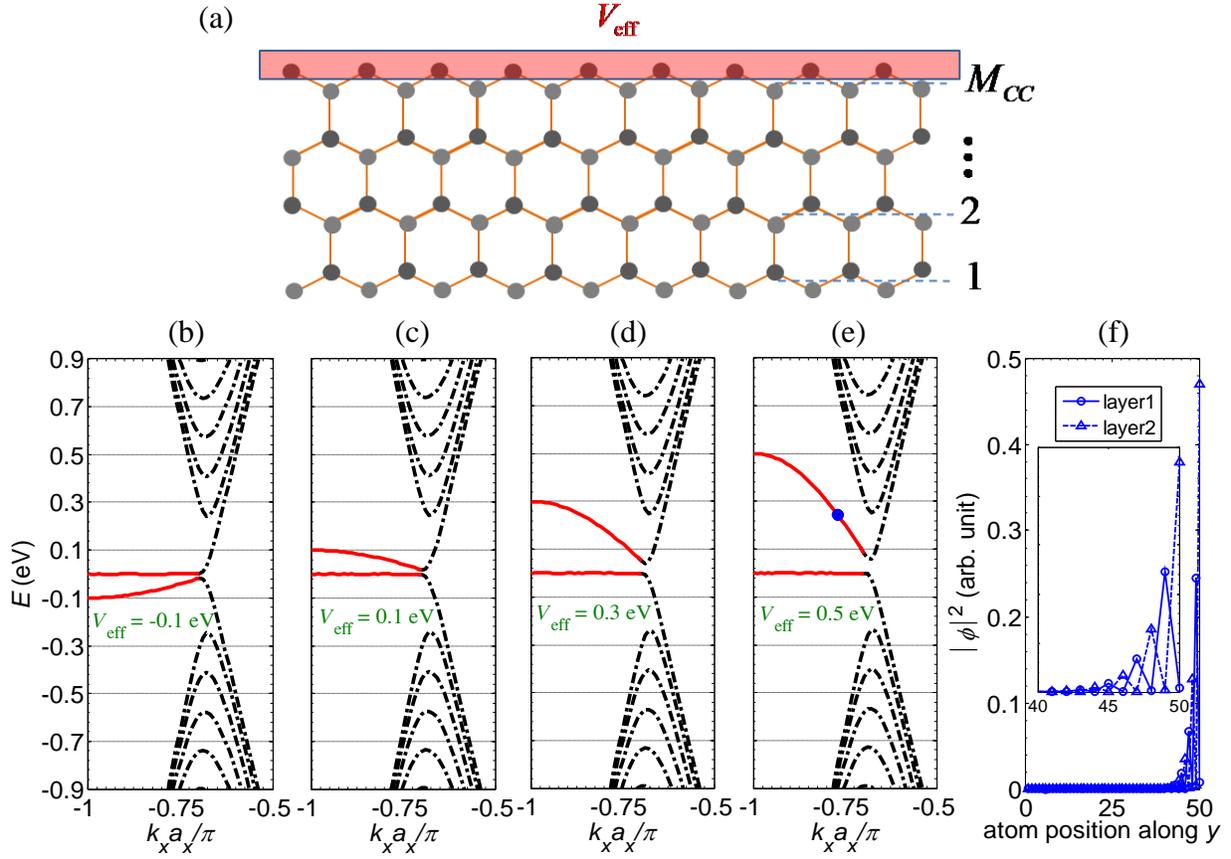

Fig 6. (a) Model of effective potential $V_{eff}$ applied on the edge atoms of a zigzag graphene ribbons to mimic the effects of G/BN interface. (b-e) Energy bands for different effective potentials applied. (f) Square wave function corresponding to the state marked with a blue dot in (e). Inset: magnification of the interface region.

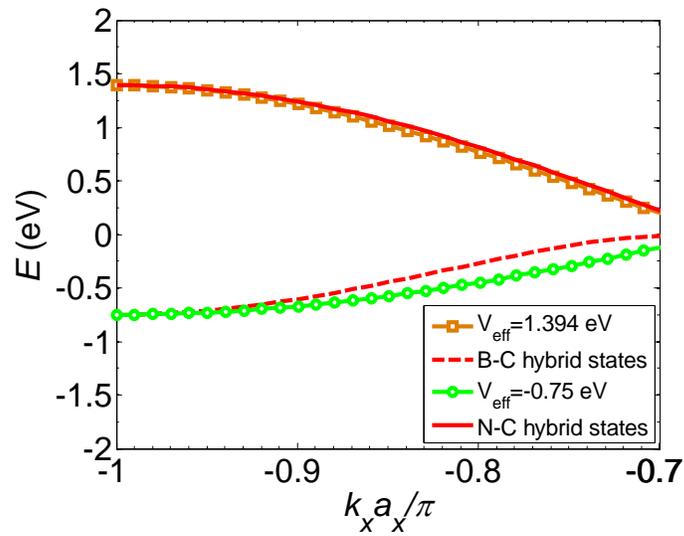

Fig 7. The new edge states in the effective potential model (symbols) compared with actual hybrid states (solid and dashed lines) in the B-C--C-N structure.

22